\documentclass[aps,prl,twocolumn,superscriptaddress, showpacs]{revtex4}

\usepackage{amsmath}
\usepackage{graphicx}
\usepackage[section]{placeins}
\usepackage{color}
\usepackage{float}
\usepackage{floatflt}
\usepackage{MnSymbol}
\usepackage{wasysym}
\usepackage[utf8]{inputenc}
\usepackage{array}
\usepackage{multirow}
\usepackage[svgnames]{xcolor}
\usepackage{titlesec}

\titleformat
   {\section}
   [block]
   {\centering\large\bfseries}
   {\thesection}
   {\baselineskip}
   {}

\begin{document}

\title{Determination of the entropy production during glass
transition: theory and experiment}

\author{H. Jabraoui} 
\affiliation{Laboratoire Physique et Chimie Th\'eoriques, CNRS - Universit\'e de Lorraine, F-54000 Nancy France.}

\author{S. Ouaskit}
\affiliation{Laboratoire physique de la matière condens\'ee, Facult\'e des sciences Ben M’sik, Universit\'e  Hassan II de Casablanca, Morocco}

\author{J. Richard}
\affiliation{Institut N\'EEL, CNRS, 25 avenue des Martyrs, F-38042 Grenoble France}
\affiliation{Univ. Grenoble Alpes, Institut N\'EEL, F-38042 Grenoble France}

\author{J.-L. Garden\footnote{Corresponding author : jean-luc.garden@neel.cnrs.fr}}
\affiliation{Institut N\'EEL, CNRS, 25 avenue des Martyrs, F-38042 Grenoble France}
\affiliation{Univ. Grenoble Alpes, Institut N\'EEL, F-38042 Grenoble France}

\date{\today}

\begin{abstract}
 A glass is a non-equilibrium thermodynamic state whose physical properties
depend on time. Glass formation from the melt, as well as the inverse process of liquid structural recovery from the glass are non-equilibrium processes. A positive amount of entropy is produced during such irreversible processes. In this paper, we address the issue of the determination of entropy production during glass transition. Firstly, we theoretically determine the entropy production by means of the statistical model of a two-level system coupled to a master equation driving the time dependency of the occupancy probability of each state. Thermodynamic cycles of the type liquid-glass-liquid are considered in order to test the validity of the Clausius theorem. Secondly, we determine experimentally the production of entropy from differential scanning calorimetry experiments on the PolyVinylAcetate glass-former. Aging experiments are also considered. From the data treatments proposed here, we are able to determine the rate of production of entropy in each part of the experiments. Although being on the order of few \% or less of the configurational entropy involved in the glass formation, the positive production of entropy is clearly determined. For all the thermodynamic cycles considered in these calorimetric experiments, the Clausius theorem is fulfilled. 
\end{abstract}

\pacs{à trouver}

\maketitle 

\section{Introduction}

There are different formulations of the second law of thermodynamics. One of them states that the infinitesimal entropy change of a system following a thermodynamic transformation, $dS$, may be written as the sum of two contributions, $dS=d_{e}S+d_{i}S$ \cite{prigo4}. $d_{e}S=\frac{\delta Q}{T}$ is the part of the entropy dealing with the heat exchange between the system and its surroundings. $d_{i}S$ is the infinitesimal internal entropy production contribution, which is produced in the system itself. This contribution is always positive, $d_{i}S\geq0$, for irreversible transformations (whatever the direction of the transformation), while it is equal to zero for reversible transformations. Since this latter contribution is always positive, the second law may be written under the form of an inequality, $dS\geq\frac{\delta Q}{T}$(Clausius inequality).
\\
Since a glass is by nature a system outside equilibrium whose properties depend on time, we can wonder what is the origin of this entropy production
term during glass formation, or during structural recovery of a super-cooled
liquid from a glassy state. What is the order of magnitude of this
contribution? Is it even experimentally measurable (by calorimetry for
example)? These questions have been tackled by Davies and Jones in their seminal paper on the thermodynamic of the glass transition \cite{davies}. Owing to the two thermodynamic variables, affinity and fictive temperature, they
estimated the production of entropy by calculation of surfaces in
the enthalpy-temperature diagram. This allowed them to estimate the
error due to irreversible processes on the measured zero-point entropy
of glasses, which is less than 2\% in the case of glycerol. As the question of the real value of the zero point entropy of glasses was of interest, Bestul and Chang discussed the applicability of the second law of thermodynamics on heating and cooling of glasses \cite{bestul}. The second law gives the upper and lower bounds, measurable by calorimetry, of the true value of the entropy at 0 K \cite{bestul}. Goldstein discussed later, in an long appendix of a paper on the excess entropy of glasses, the reliability of the calorimetric method \cite{goldstein2}. He also investigated the difference between the real value of the entropy ($\varDelta S$) and that measured by calorimetry ($\varDelta S_{app}$). Based on Davies and Jones paper, he proposed to use the fictive temperature to discuss the problem, and he proposed several simulations \cite{goldstein2}. The conclusion is that the effect of the irreversibility of the transformations on calorimetric measurements is negligible. More recently, Möller and co-workers used the framework of non-equilibrium thermodynamics to simulate the temperature behavior of the production of entropy during cooling and heating of a glass \cite{gutzow3}. Tropin and colleagues used the same model in order to simulate the temperature behavior of the production of entropy over decades of different heating and cooling rates \cite{schmelzer1}. They even considered cyclic processes from differential scanning calorimetry (DSC) experiments on polystyrene and PLC to verify the Clausius theorem \cite{schmelzer1}. Moreover, Tombari and Johari discussed
on the validity of the Clausius theorem during glass transition \cite{tombari2}. Based on highly sensitive calorimetric experiments carried out with a home-made calorimeter, they showed by means of an accurate analysis of the uncertainty in the calorimetric measurement that the production of entropy is too small to demonstrate
whether the Clausius theorem is likely to hold or not, and consequently that it is to small to invalidate the calorimetric recording of entropy \cite{tombari2}. 
  \\
In the present work, we also discuss on the determination of the production of entropy during glass transition. In a first part, we address the question on a theoretical point of view. The master equation approach to the non-equilibrium glass transition developed by Bisquert from a two-level system (TLS) is used for that purpose \cite{bisquert1}. In a second part, the issue is experimentally addressed. We describe a procedure in order to extract the production of entropy generated by a glass-former (here the PolyVinylAcetate (PVAc)) during cooling and heating scans of a differential scanning calorimeter (DSC). We also address this issue for thermodynamic cycles including annealing stages.

\section{Master equation approach of the non-equilibrium glass transition}

The TLS model has long been used to reproduce experimental results in glass research and in particular the memory effect \cite{aquino}. For example, it successfully describes the low-temperature properties of glasses, \textit{e.g.}, the linear temperature dependence of the specific heat (see references in \cite{aquino}). Takada and colleagues start from a TLS to deeply discuss the notion of configurational entropy, residual entropy of glass on a statistical point of view, addressing problems close to such as discussed here \cite{takada}.  From a macroscopic point of view, Bisquert started from a TLS coupled to a master equation with the purpose to simulate basic thermodynamic features of the glass transition \cite{bisquert1}. In particular, by means of this approach, it is possible to simulate classical properties of glasses. For example, during cooling it is possible to observe the saturation to constant values of the energy and entropy due to the freezing of the system configurations, while the averaged energy and entropy of the equilibrium TLS continue to decrease. It is also possible to recover the behavior of the specific heat as experimentally measured during a glass transition. A positive enthalpy peak is observable on heating due to the enthalpy recovery process during unfreezing. It is also possible to observe the effect of negative overshoot of the specific heat (sometimes called the "negative specific heat effect") at the beginning of heating. The so-called fictive temperature associated to the slow modes of the glass is also perfectly calculable with this model \cite{bisquert1}. 
\\  
In Fig.(1) a TLS is represented with two energy minima, the ground state called state $0$ at energy $\epsilon_{0}=0$,
and the excited state called state $1$ at energy $\epsilon_{1}=\epsilon$.
\begin{figure} 
\begin{center}
 \includegraphics[width=8.5cm]{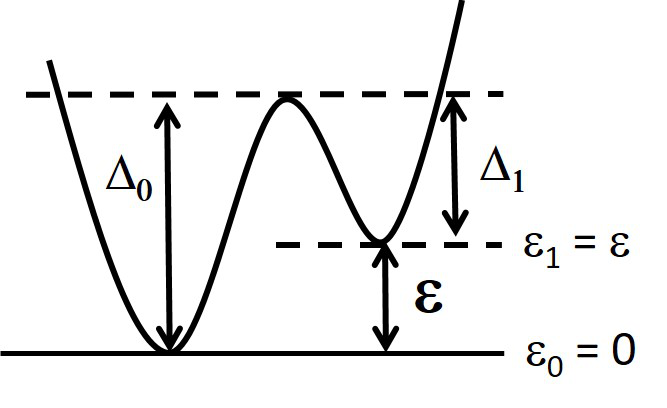}
 \end{center}
 \caption{Double-well potential of a two-level-system.}
 \label{Fig1}
\end{figure}
The value of the energy barrier is $\triangle_{0}=1$ in units of
energy. From this TLS it is possible to calculate the occupancy probability
of each state at equilibrium with $p_{0e}=1/Z$ and $p_{1e}=e^{-\epsilon/T}/Z$
with the partition function $Z=1+e^{-\epsilon/T}$. From the probabilities,
we have access to all the macroscopic properties in equilibrium, with
in particular the energy of the TLS, $E_{eq}=\left\langle \epsilon_{i}\right\rangle =\epsilon p_{1e}$.
Having the mean energy of the system in equilibrium, the heat capacity
of the TLS is $C_{eq}=dE_{eq}/dT=\varepsilon^{2}e^{\epsilon/T}/T^{2}\left(1+e^{\epsilon/T}\right)^{2}$,
which presents a Schottky anomaly in the 'low-temperature' regime.
The kinetic of the energy distribution in the TLS can be represented
by a master equation driving the time dependency of the occupancy
probability of each state. In focusing on the excited state, we have
$dp_{1}/dt=\omega_{0,1}p_{0}-\omega_{1,0}p_{1}$ where $\omega_{0,1}$and
$\omega_{1,0}$ are the transition rates from the levels 0 to 1 and
1 to 0 respectively. It is assumed that the two transition rates are
of the type $\omega_{0,1}=\nu_{0}e^{-\triangle_{0}/T}$ and $\omega_{1,0}=\nu_{0}e^{-\triangle_{1}/T}$.
This would say that particles have to overcome the height barriers
$\triangle_{0}$ and $\triangle_{1}$ in order to be in
the minimum 0 or 1. Such an assumption amount
to consider a mean relaxation time of Arrhenian type. Indeed, using
the principle of detailed balance occurring in equilibrium $p_{0e}/p_{1e}=\omega_{1,0}/\omega_{0,1}$,
the master equation is rewritten under the form of a simple first-order differential equation for the occupancy probability to be in
the state of energy level 1: 
\begin{equation}
\frac{dp_{1}}{dt}=-\left(\frac{p_{1}-p_{1e}}{\tau}\right)
\end{equation}
with $\tau=1/(\omega_{0,1}+\omega_{1,0})$ is the mean kinetic relaxation
time of the TLS system when submitted to a perturbation. Here the
perturbation is the temperature change driving the $p_{1e}$term and
$\tau$. In defining a linear time dependent law for temperature in
cooling/heating $T=T_{0}\pm bt$ , Eq.(1) is numerically solved. Once $p_{1}$
is calculated as a function of time/temperature, the configurational
energy $E=\left\langle \epsilon_{i}\right\rangle =\epsilon p_{1}$,
the configurational entropy $S=-k_{B}\left\langle lnp_{i}\right\rangle =-k_{B}\left(p_{0}lnp_{0}+p_{1}lnp_{1}\right)$,
and the configurational heat capacity are calculated along temperature
ramps. In Fig.(2), the energy of a TLS is represented in equilibrium
(black line) and during cooling and heating (blue and red lines) for temperature rates
$b=10^{-5}$(rapid cooling and heating) and $b=10^{-7}$(slow cooling
and heating) in unit of Kelvin per second. In the rest of the paper, all the parameters shown in the different figures, whose temperature behavior is simulated from the model, are in arbitrary units.
\begin{figure} 
\begin{center}
\includegraphics[width=8.5cm]{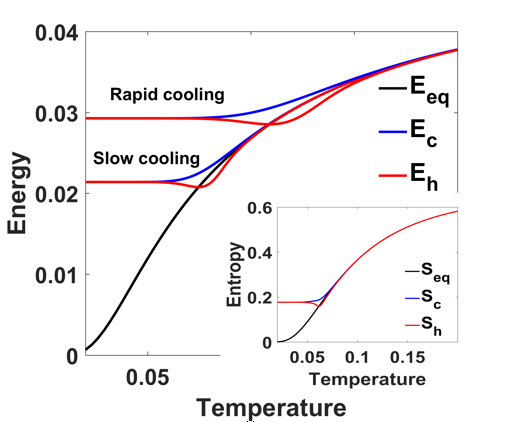}
 \end{center}
\caption{Energy of the TLS as a function of temperature during a rapid cooling and a slow cooling, and during the successive heating. The energy of a TLS remaining at thermodynamic equilibrium during cooling and heating is also shown. \textit{Inset}: Entropy of a TLS during cooling and heating as well as entropy of a TLS remaining at equilibrium.}
 \label{Fig2}
\end{figure}
The blue lines correspond
to energies during cooling, and red lines correspond to energies during
heating succeeding the previous cooling. During cooling at a given
rate, the energy starts to depart from its equilibrium value (progressive
freezing) and becomes constant when the system is completely vitrified
(glassy state). The lower the cooling temperature rate is, the lower the energy at which the system starts to depart from equilibrium is (smaller value of $T_{g}$). During heating
the energy cross the equilibrium line due to a damping effect, and it returns
to the equilibrium line from below. The TLS energy behaves like a glass configurational
energy. In using the statistical definition of entropy above, we observe
that the configurational entropy of the TLS is approaching a constant
value during cooling, and that it behaves exactly like the energy during cooling and
successive heating (see inset in Fig.(2) for the rate $b=10^{-7}$). The constant value of this configurational entropy at low temperature is the so-called zero-point entropy of a glass, while
it is equal to zero for the equilibrium state. In Fig.(3), the configurational
heat capacities belonging to the respective TLS energies are represented
as a function of temperature ($C=dE/dT=\epsilon dp_{1}/dT$). 
\begin{figure} 
\begin{center}
 \includegraphics[width=8.5cm]{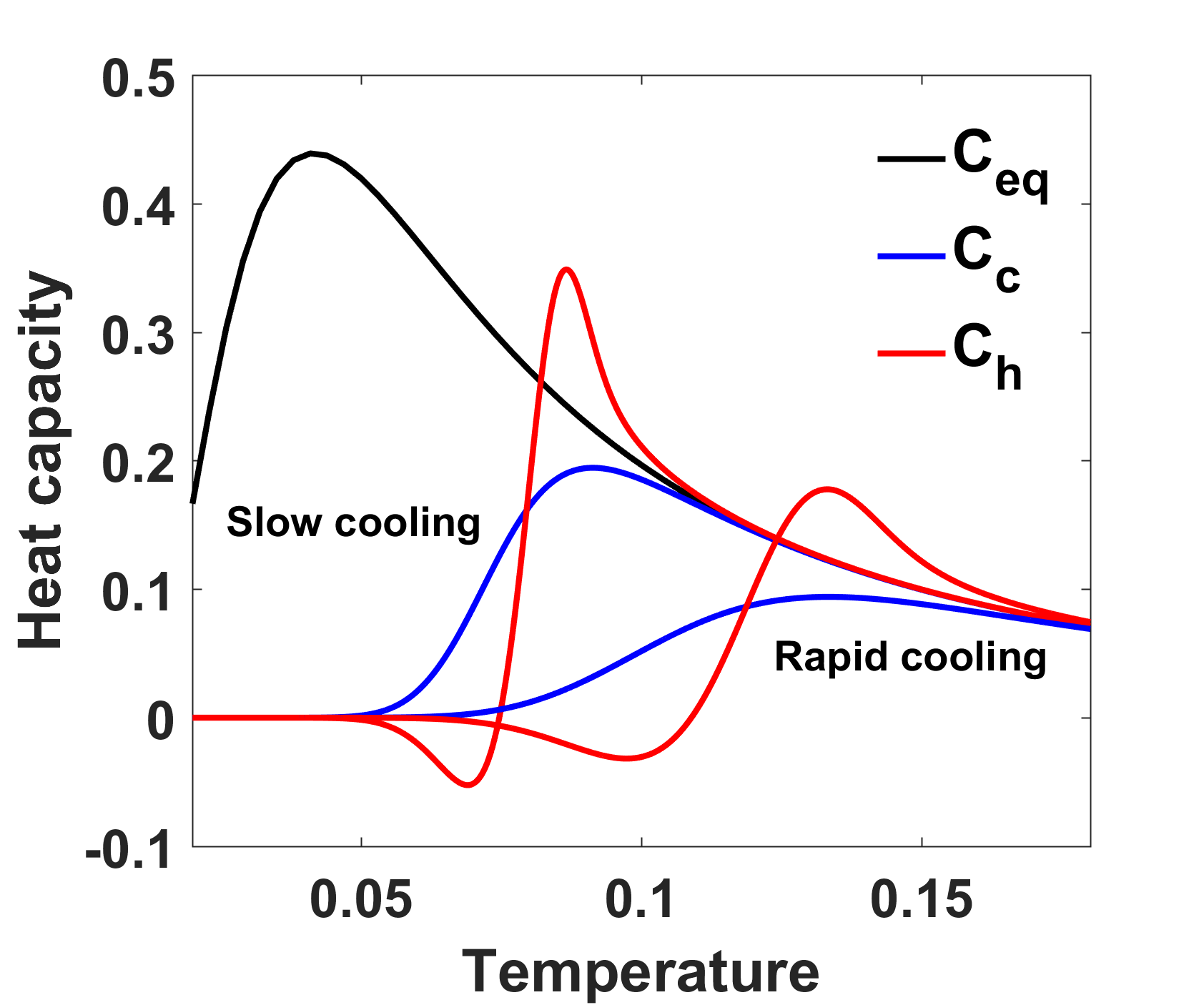}
 \end{center}
 \caption{Configurational heat capacity of a TLS as a function of temperature during a rapid cooling and a slow cooling, and during the successive heating. The configurational heat capacity of a TLS remaining at thermodynamic equilibrium during cooling and heating showing a Schottky anomaly.}
 \label{Fig3}
\end{figure}
Heat capacity drops during cooling, that are typical of a glass transition,
are observed at different glass temperatures depending on the temperature
rates. At the beginning of heating the small decrease in the heat
capacity (negative heat capacity effect) is observed as well as the
enthalpy recovery peaks at higher temperature, both effects being typical
of glass-formers. Magnitudes and positions of these peaks are temperature
rate dependent. Due to intrinsic dynamics of TLS, it is thus possible
to kinetically avoid a Schottky anomaly when temperature cooling rate is
too fast. Tool defined the fictive temperature of a glass in assuming
that a property $P$ (energy, volume, index of refraction, etc...)
of a non-equilibrium glass at temperature $T$ is equal to the property $P_{eq}$ of the same system at equilibrium at the fictive temperature $T_{f}$, $P(T) = P_{eq}(T_{f})$ \cite{tool1,tool2}. When the system
is completely glassy, the fictive temperature is constant, and written
$T_{f}^{'}$, the limiting fictive temperature. This is why this temperature
may be associated with the specific temperature of the slow configurational
modes in a glassy system \cite{tool1,tool2}. Here the fictive temperature
is calculated by means of the TLS configurational energy:
\begin{equation}
E(T)=E_{eq}(T_{f})
\end{equation}
Once $p_{1}(T)$ has been calculated from Eq.(1), $T_{f}$ is calculated as a function of temperature at any time solving the equation $\epsilon p_{1}(T)=\epsilon e^{-\epsilon/T_{f}}/(1+e^{-\epsilon/T_{f}})$. The temperature
dependency of the fictive temperature, not represented here, is similar
to the energy and entropy behaviors. By means of this approach, we would like now to go further by the calculation of the entropy production rate during glass transition.

\section{Thermal model of the glass transition} 
In a previous paper, some of the present authors have established correspondence formula between the fictive temperature framework and non-equilibrium thermodynamic framework of de Donder and Prigogine \cite{garden1}. Actually, the existence of a glass fictive temperature proposed
by Tool is equivalent to consider that the glass-system is composed of two sub-parts (see Fig. (4)). 
\begin{figure} 
\begin{center}
 \includegraphics[width=4.25cm]{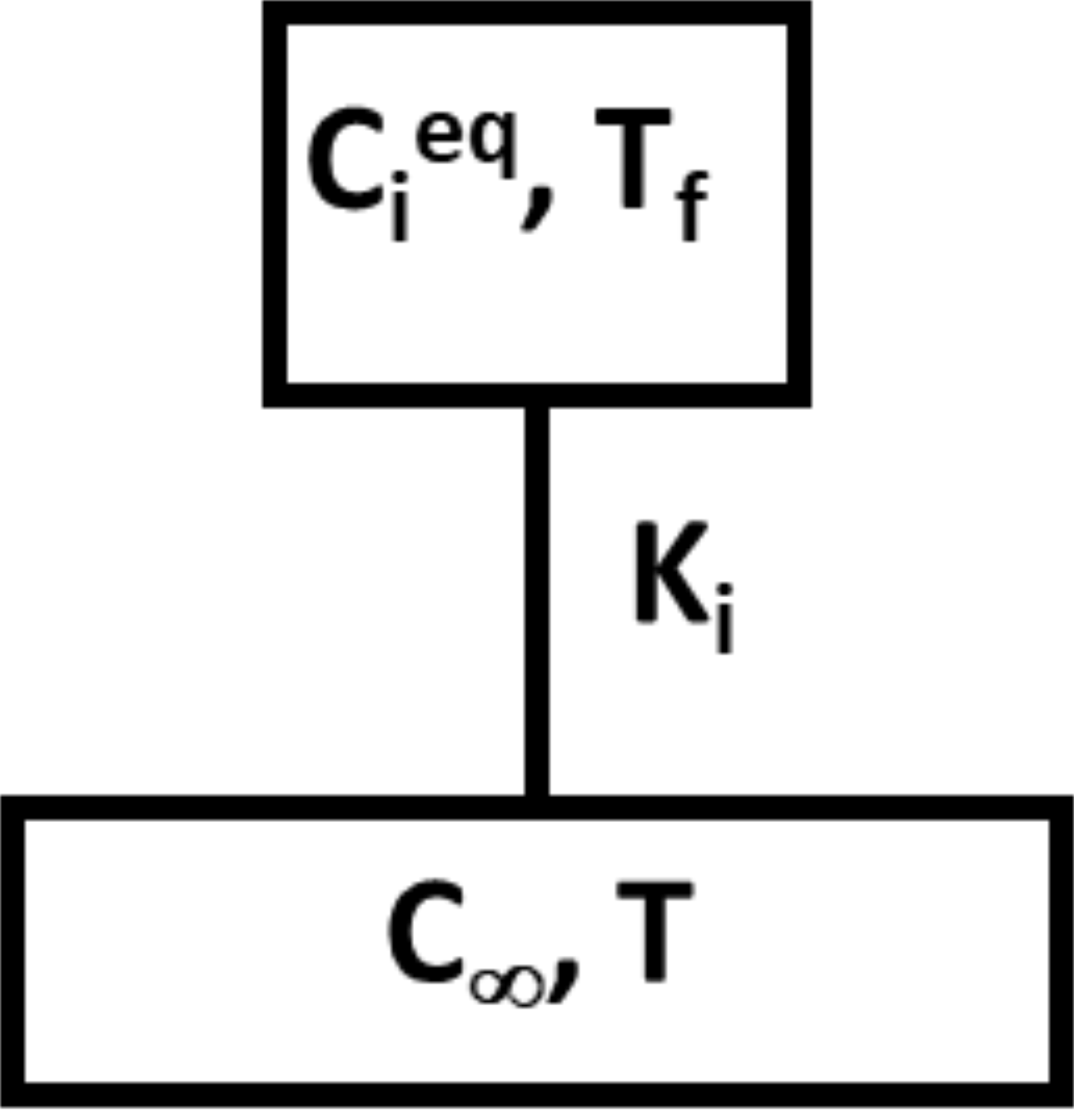}
 \end{center}
 \caption{Thermal model of a glass from a double-system. The first system is at the equilibrium temperature T. The other system is thermally coupled to the first one by a thermal conductance $K_{i}$. It has a fictive temperature $T_{f}$ and an equilibrium configurational heat capacity $C_{i}^{eq}$.}
 \label{Fig4}
\end{figure}
The first one, of temperature $T$,
is the part of the system composed by the 'fast' internal degrees
of freedom (generally vibrations or phonon-bath). This part, of the heat
capacity, $C_{\infty}$, plays the role of a thermal bath for the second
part, composed by the 'slow' configurational modes (it may be rotations,
molecular changes, translations,...). Here, 'fast' and 'slow' means that some degrees of freedom reach equilibrium more rapidly than other ones when the system is submitted to a perturbation (temperature change). This second part has a temperature
$T_{f}$ and a heat capacity $C_{i}^{eq}$. At equilibrium, the thermal
bath imposes the temperature of the configurational part and $T_{f}=T$
at each instant. But, during a dynamic event, induced by cooling or
heating, the presence of an 'internal thermal conductance' $K_{i}$ generates a delay in
the response of the slow modes with respect to the fast ones that have reached equilibrium instantaneously. This $K_{i}$ parameter defines a
mean relaxation time $\tau=C_{i}^{eq}/K_{i}$. What does this internal thermal conductance physically mean at molecular
scale is outside our considerations in this paper. Now, let
us assume that the TLS is equivalent to this two-body thermal
system. In this case, the configurational energy, the configurational
entropy and the configurational heat capacity of the TLS are equivalent to the mean energy,
the mean entropy and the heat capacity of the whole system, less
the energy, the entropy and the heat capacity of the phonon-bath sub-system.
Consequently, the configurational heat capacity of the TLS is: 
\begin{equation}
C_{TLS}=C_{i}=C_{tot}-C_{\infty}
\end{equation}
The energy balance applied to the two-body system is:
\begin{equation}
C_{i}^{eq}\frac{dT_{f}}{dt}=C_{i}\frac{dT}{dt}=K_{i}(T-T_{f})
\end{equation}
The first equality comes from the temperature derivative of the Tool
Eq.(2), $dT_{f}/dT=C_{i}(T)/C_{i}^{eq}(T_{f})$. The second equality
of Eq.(4) is the amount of heat flowing between the two sub-parts
of the system at different temperatures. As shown in Fig.(3), in the high temperatures range there is $C_{TLS}=C_{i}^{eq}$. The system is at equilibrium with the equilibrium value of the TLS configurational heat capacity, and we have $T_{f}=T$. This equilibrium TLS configurational heat capacity corresponds to the super-cooled liquid heat capacity less the glass heat capacity of a glass-former. In the low temperatures range, there is $C_{TLS}=C_{i}=0$ (since $C_{tot}=C_{\infty})$, and $T_{f}=T'_{f}$. The system is frozen in one glassy state with a configurational heat capacity equals to zero. 

\section{Entropy production rate}
From the thermal model above, the production of entropy may be defined like in the practical case of an irreversible heat-flow between two
bodies at different temperatures \cite{garden1}: 
\begin{equation}
\sigma_{i}=\frac{d_{i}S}{dt}=\left(\frac{1}{T_{f}}-\frac{1}{T}\right)\times K_{i}(T-T_{f})=\left(\frac{1}{T_{f}}-\frac{1}{T}\right)\times C_{i}\frac{dT}{dt}
\end{equation}
We observe that this entropy production rate is written as a product of
a thermodynamic force (the perturbation) with a thermodynamic flux (the response)
\cite{garden1}. The thermodynamic force is the difference of the inverse of temperatures of the bodies, and the thermodynamic
flux is the corresponding heat flux between these two bodies. Such an expression has also been derived by Davies and Jones with non-equilibrium thermodynamic framework and Tool fictive temperature concept \cite{davies}, and used by Goldstein to provide calculations of errors in entropy measurement with calorimetry \cite{goldstein2}. Considerations on the generalization of the entropy function for non-equilibrium systems undergoing irreversible processes have also led Bridgman to derive such an equation for the entropy production contribution \cite{bridgman}. Since
all the parameters in Eq.(5) are calculable with the TLS model, we
can simulate the production of entropy during cooling and heating
(see Fig. (5)). 
\begin{figure} 
\begin{center}
 \includegraphics[width=8.5cm]{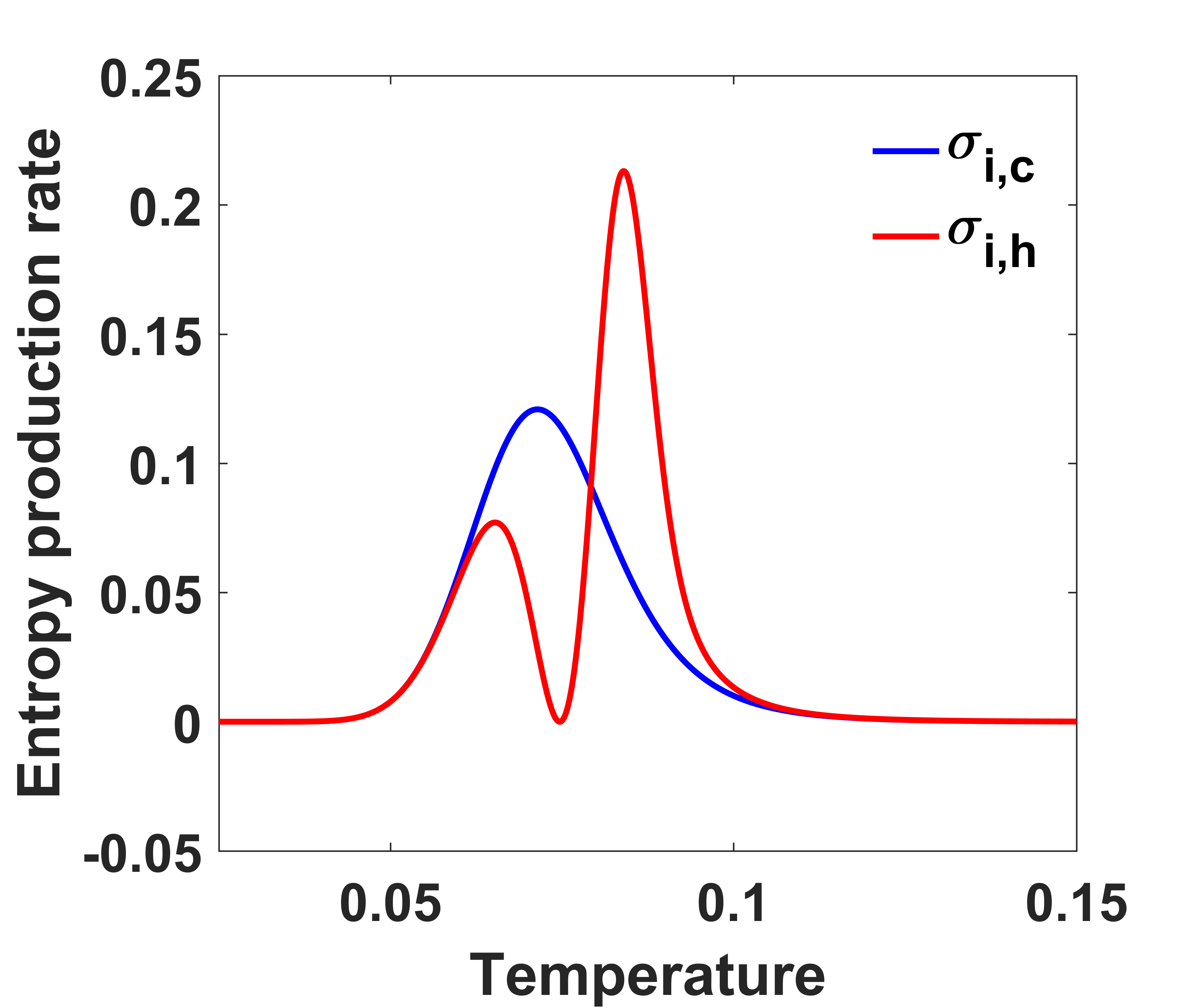}
 \end{center}
 \caption{Entropy production rate of a TLS as a function of temperature during cooling and the successive heating.}
 \label{Fig5}
\end{figure}
The entropy production rate is shown for a temperature rate of $b=10^{-7}$ during cooling and heating. There is a broad peak of entropy production during
cooling, and two peaks during the successive heating. The result of
such simulations is similar to that obtained by Möller and colleagues,
demonstrating another time that this approach is equivalent to the
classical non-equilibrium thermodynamics \cite{gutzow3} (see also Ref. \cite{garden6}
for more details). Now that we have at our disposal an expression for the
production of entropy, which is calculable, let us express the configurational
entropy change of the system as a function of the same parameters. From the second law of thermodynamics, the total entropy change of the system is written:
\begin{equation}
dS_{tot}=d_{e}S+d_{i}S
\end{equation}
where $d_{e}S=\delta Q/T$ is the entropy exchange (due to heat exchange) between the system and the surroundings, and $d_{i}S$ is the infinitesimal production of entropy that may be obtained from Eq.(5) above. In the case of the TLS, the heat exchange with the surroundings is the heat exchange between the two bodies (since the surroundings is the thermal bath), and it is equivalent to the mean configurational energy change $d\left\langle E\right\rangle =C_{i}dT$ (since there is no work involved), and we have for the total configurational entropy variation:
\begin{equation}
dS_{tot}=\frac{C_{i}}{T}dT+\frac{(T-T_{f})}{T_{f}}\frac{C_{i}}{T}dT
\end{equation}
We notice by the way that the production of entropy contribution (second term of the right hand side term in Eq.(7)) is just the entropy due to configurational energy exchanges multiplied by a kind of Carnot factor $(T-T_{f})/T_{f}$. It could be of interest to express now the second law of thermodynamics under the following equality, $dS_{tot}/d_{i}S=1+\Delta T/T_{f}$. In summing the two contributions in Eq. (7) above, we come to an important conclusion
that the total configurational entropy change in a system departed from equilibrium
is obtained by integration of the configurational
heat capacity divided by the fictive temperature, and not by the classical
temperature as it is generally derived from calorimetric experiment:
\begin{equation}
dS_{tot}=\frac{C_{i}}{T_{f}}dT
\end{equation}
In integrating this expression, we may verify that we
obtain exactly the same curve than those shown in inset in Fig.(2) from the classical expression $S=-k_{B}\left\langle lnp_{i}\right\rangle =-k_{B}\left(p_{0}lnp_{0}+p_{1}lnp_{1}\right)$, confirming that the two expressions for the total configurational entropy are identical:
\begin{equation}
\int dS_{tot}=\int\frac{C_{i}}{T_{f}}dT=-k_{B}\left\langle lnp_{i}\right\rangle
\end{equation}
The expression (7) or (8) above means that, if we consider only the entropy variation due to the configurational energy exchange between the two parts of the double system (first term in the RHS of Eq.(7)), $\int C_{i}/TdT$, we miss the positive production of entropy contribution,
whatever the irreversible path followed by the system. This can be
simply verified in presenting on the same graph in Fig.(6) $C_{i}/T$
and $C_{i}/T_{f}$ for the cooling and the successive heating (temperature rate of $b=10^{-5}$). 
\begin{figure} 
\begin{center}
 \includegraphics[width=8.5cm]{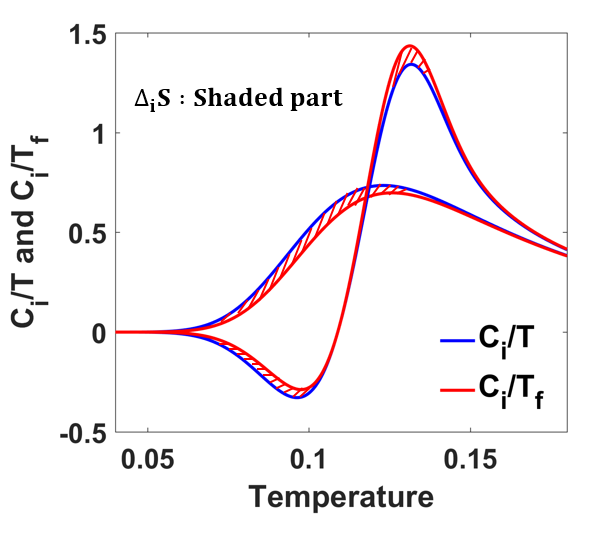}
 \end{center}
 \caption{Configurational heat capacity of a TLS divided by the temperature, and configurational heat capacity of a TLS divided by the fictive temperature, as a function of temperature during cooling and the successive heating.}
 \label{Fig6}
\end{figure}
On cooling, the $C_{i}/T_{f}$ curve
is below the $C_{i}/T$ curve during the glass transition range. Since
the temperature is decreasing ($dT/dt<0$), that is to say that the
production of entropy is positive, $\sigma_{i,c}=(C_{i}/T_{f}-C_{i}/T)\times dT/dt>0$.
On heating, there are two distinct temperature ranges where the $C_{i}/T_{f}$
curve is above the $C_{i}/T$ curve. Since on heating $dT/dt>0$,
that is to say that the production of entropy is also positive on
these two ranges, $\sigma_{i,h}=(C_{i}/T_{f}-C_{i}/T)\times dT/dt>0$.
The difference between the two curves in Fig.(6)(shaded part)is exactly what is presented in Fig.(5), with one positive peak under
cooling and two positive peaks on heating. 

\section{Clausius theorem}
The integration of Eq.(7) above on one thermodynamic cycle yields directly to the Clausius theorem:
\begin{equation}
\oint\frac{C_{i}}{T}dT=\oint\frac{\delta Q}{T}=-\oint d_{i}S<0
\end{equation}
The integral on one thermodynamic cycle of the heat exchange (divided by the temperature) between a system and its thermal surrounding is always negative. It is equal to zero for a reversible transformation along the cycle. We have used the fact that entropy is a state function and that $\oint dS_{tot}=0$. In the case of the glass transition, considering the liquid-glass-liquid cycle the theorem must be fulfilled since this transformation is by essence irreversible and that entropy is produced whatever the thermal history. In the graph in Fig.(7), we have integrated the $C_{i}/T_{f}$ on the cycle (red
solid line) as well as the $C_{i}/T$ (blue dashed line)
when starting from the same value of entropy at high temperature in
the equilibrium liquid ($T=0.3$ in our simulations).
\begin{figure*} 
\begin{center}
\includegraphics[width=18 cm]{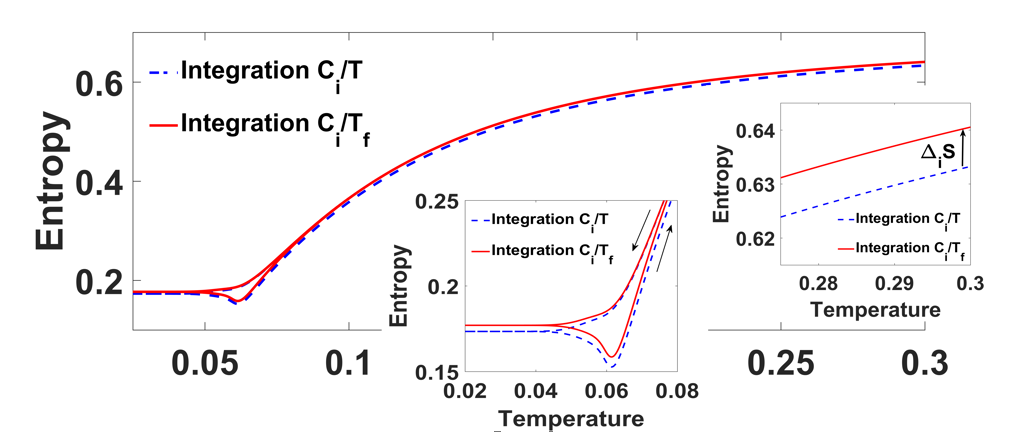}
\end{center}
\caption{Configurational entropy of a TLS obtained by integration of $C_{i}/T$ and $C_{i}/T_{f}$ as a function of temperature during cooling and the successive heating. \textit{Inset1}: magnification of the configurational entropy in the low temperature range, where we observe that the integration of the $C_{i}/T$ gives an entropy remaining below that calculated from the integration of the $C_{i}/T_{f}$. \textit{Inset2}: magnification of the configurational entropy in the starting and final temperature range where it is observed that the integration of the $C_{i}/T$ gives a final value of the entropy which is distinctly below the final value of the entropy calculated from the integration of the $C_{i}/T_{f}$, this latter configurational entropy being identical to the starting value.
}
 \label{Fig7}
\end{figure*}
Although the integration of $C_{i}/T_{f}$ on the cycle brings the entropy back
to this same initial value, the integration of the $C_{i}/T$ brings
the entropy toward a lower value (see Inset2 in Fig.(7)). The difference
between both values are the net positive entropy produced all along the
irreversible path followed by the system. This contribution is missing
in the integration of $\delta Q/T$. Since there is a positive
production of entropy during cooling, we observe a lower
value of the zero-point entropy of glass when only $\delta Q/T$ is
considered (see Inset1 in Fig.(7)). During heating, the dashed line continues to depart from the solid line due to the existence of two temperature ranges where entropy is produced. 
\\
As a conclusion, the TLS model coupled to a master equation, not only accounts for the basic features of the glass transition, but also obeys the Clausius theorem.
The question now arises to know whether these considerations may be experimentally checked from calorimetric experiments on glasses.

\section{Experiments}
\subsection{Material and method}
The calorimeter used for differential scanning calorimetry (DSC) experiments is a commercial apparatus microDSC-III of Setaram company. This is a heat-flux DSC with two hastelloy batch cells of empty volume of 1 mL. The temperature is controlled either during heating or during cooling with scanning rates ranging between about 0.1 K/min to 1.2 K/min. The working temperature ranges between about -15 °C to 115 °C. 
\\
The sample used for the experiments is a model of polymeric glass-former, the PolyVinylAcetate [PVAc, ($C_4H_6O_2)_n$]. The average molecular weight is 157 kg/mol and the polydispersity index (PDI), which indicates the distribution of individual
molecular masses in a polymer, is 2.73. The PVAc is first conditioned during 20 h in
an oven under vacuum at a temperature of 150 °C, well above
the glass transition temperature, in order to remove all the water and bubbles from the melt. The mass is measured (m = 148.36 mg) and the sample is sealed in the sample-cell of the calorimeter.
\\
The reference material is a polyimide from DuPont company (Vespel) having not any transitions in the considered temperature range. It is machined in order to have a bar of 0.3 mm in diameter and of 5.34 mm long (m = 214.6 mg). Before using this material as a neutral reference for PVAc measurements, it is used as a calibration material for calibrating the DSC. In this preliminary step, the Vespel is used as a sample, and the reference cell is let empty. Many experiments are carried out at different scanning rates, either during heating or during cooling. This initial procedure allows us to correct, up to a certain extent, the DSC inaccuracy as a function of the scanning rate, and the DSC asymmetry between heating and cooling for each scanning rate.

\subsection{Specific heat measurements}
Five different experiments have been considered. The exact protocol of each experiment is summarized in table I. 

\begin{table*}
\begin{tabular}{|c|c|c|c|c|c|c|c|c|c|c|}
\hline 
 & Exp. 1  & Exp. 1 & Exp. 2 & Exp. 2 & Exp. 3 & Exp. 3 & Exp. 4 & Exp. 4 & Exp. 5 & Exp. 5\tabularnewline
\hline 
 & cooling & heating & cooling & heating & cooling & heating & cooling & heating & cooling & heating\tabularnewline
\hline 
$T_{high}$ (K) & 364.16 & 364.16 & 364.16 & 364.16 & 364.16 & 364.16 & 364.16 & 364.16 & 364.16 & 364.16\tabularnewline
\hline 
waiting time (min) & $>$20 & - & $>$20 & - & $>$20 & - & $>$20 & - & $>$20 & -\tabularnewline
\hline 
scanning rate (K/min) & $-$1.2  & $+$1.2  & $-$ 0.4 & $+$ 1.2 & $-$1.2  & $+$1.2 & $-$1.2  & $+$1.2 & $-$1.2  & $+$1.2\tabularnewline
\hline 
$T_{aging}$(K) & - & - & - & - & 293.3  & - & 293.3  & - & 293.3  & -\tabularnewline
\hline 
aging time (h) & - & - & - & - & 60 & - & 20 & - & 6 & -\tabularnewline
\hline 
$T_{low}$ (K) & 267.95 & 267.95 & 267.95 & 267.95 & 267.95 & 267.95 & 267.95 & 267.95 & 267.95 & 267.95\tabularnewline
\hline 
\end{tabular}
\caption{Experimental protocols}
\end{table*}

In a first step, we consider only the experiments 1, 2 and 3. In the experiments 1 and 2, the PVAc is firstly maintained at the temperature of 364.16 K during more than 20 min. At this temperature, the PVAc is in the liquid state and its memory is completely erased. Then it is cooled from this temperature to the lower temperature of 267.95 K at a scanning rate of - 1.2 K/min for the experiment 1 and - 0.4 K/min for the experiment 2. The sample is then instantaneously heated to the initial temperature in the liquid state at 1.2 K/min. During the cooling and the heating, the differential heat-flux is measured. The differential PVAc heat capacity is calculated by means of the following calorimetric equations:
\begin{equation}
\varDelta C=C_{PVAc}-C_{Vespel}=-\frac{\varDelta P_{mes}}{\beta}
\end{equation}
where $C_{PVAc}$ and $C_{Vespel}$ are the PVAc and Vespel heat capacities
of the sample and reference respectively, $\varDelta P_{mes}$ is the measured
differential heat-flux, and $\beta=dT/dt$ is the temperature
scanning rate of the considered experiment parts.  $C_{PVAc}(T)$ is obtained with the use of the $C_{Vespel}(T)$ curve derived from the previous calibration procedure of the DSC with the Vespel as a sample.
\\
In the third experiment, the sample is maintained during more than 20 min at the same initial temperature of 364.16 K and it is cooled at a scanning rate of -1.2 K/min till the temperature of 293.30 K, where it remains at this aging temperature during about 60 h. After this annealing process, it is cooled again toward the lower temperature of 267.95 K at the same scanning rate. It is then instantaneously heated from this temperature toward the initial higher temperature at a scanning rate of 1.2 K/min.
\\ 
In Fig.(8), the PVAc specific heat is shown as a function of temperature during the cooling of the experiment 1 and 2 (blue solid and dashed lines), and during heating scans at 1.2 K/min of the three experiments. 
\begin{figure} 
\begin{center}
 \includegraphics[width=8.5cm]{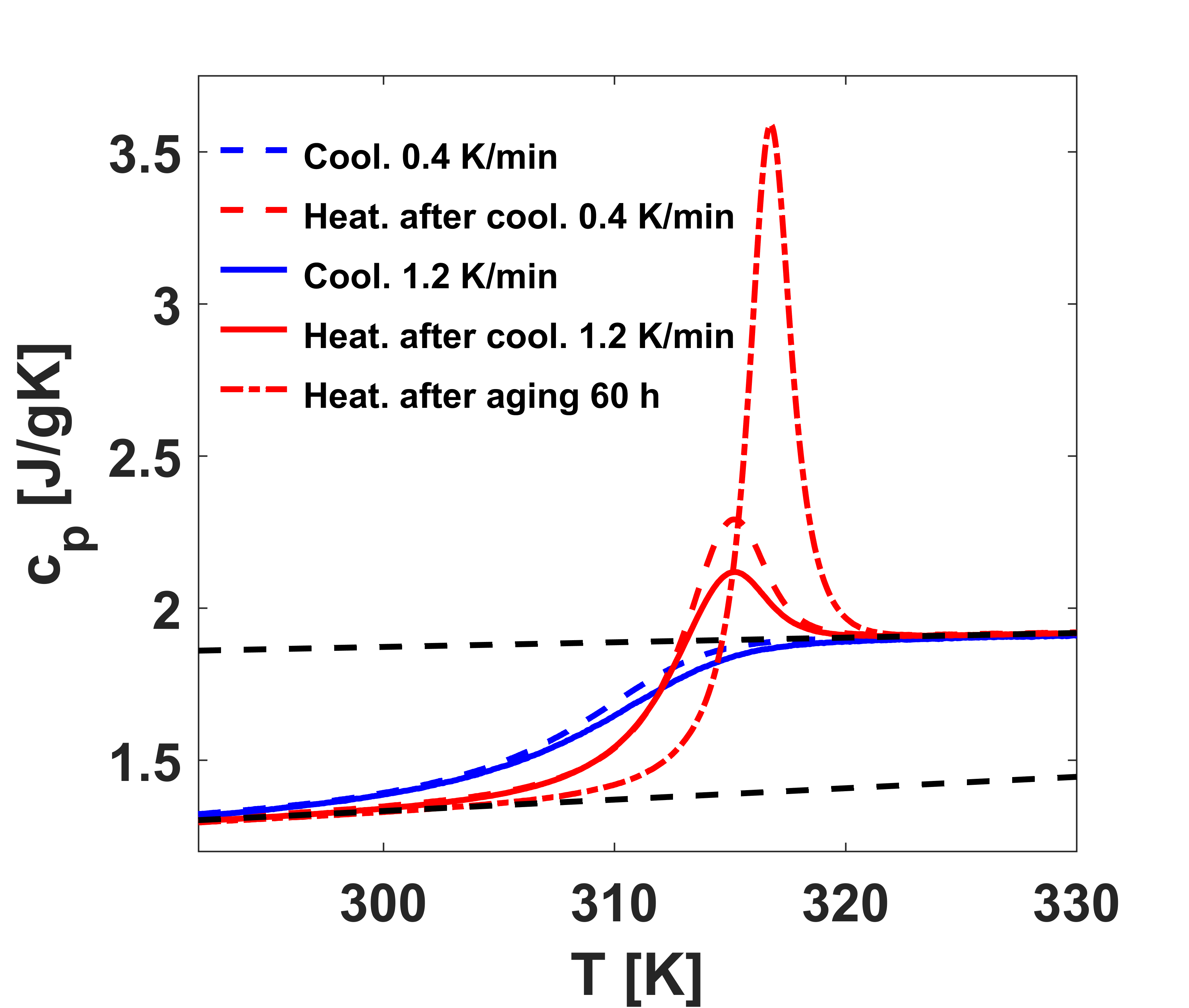}
 \end{center}
 \caption{PVAc specific heat as a function of temperature on cooling and heating for the three experiments 1, 2 and 3 such as described in table I. The liquid and glass specific heat baselines (straight lines) are shown as black dashed lines.}
 \label{Fig8}
\end{figure}
The PVAc specific heats of the experiments 1 and 2 without aging is traced with red solid and dashed line respectively, and the PVAc specific heat issued from the experiment with a previous aging is traced with red dotdashed line. At 340 K, we obtain a value of $C_{p}$ which is only 0.018 J/gK higher (around 0.9 \%) than the value obtained by Sasabe and Moynihan \cite{moy4}. In the liquid state, the slope of $C_{p}(T)$ that we obtained is 15 \% lesser than the slope measured by the same authors \cite{moy4}. The shapes of the curves are coherent with a glass transition. During cooling, the liquid is progressively transformed into a glass. This is seen as a sharp drop in the heat capacity around the glass temperature $T_{g}$. The molecular movements that are characteristic of the liquid are progressively inhibited because of the drastic exponential increase of their relaxation time. They are frozen-in and they do not participate anymore to the measurement of the heat capacity at such given scanning rates. For the cooling at the smaller cooling rate (blue dashed line), the drop is shifted toward lower temperature (slightly smaller value of $T_{g}$). During heating, we observe a heat capacity overshoot. This peak is more pronounced for experiment with aging (red dotdashed line) than for experiment without aging. It is also more pronounced for experiment 2 (red dashed line) with a smaller cooling rate than in the experiment 1 (red solid line). The area under these peaks during heating corresponds to the internal energy decrease during the sample past-history. For instance, we have to supply the same amount of energy during this heating than the energy decrease during the previous cooling in order that the system returns to its initial equilibrium liquid sate. During annealing at 293.3 K, the PVAc internal energy had much longer time to decay in its energy landscape than during a simple cooling scan without aging. This is why $c_{p}$ overshoots are generally greater after such previous annealing. The shift of this peak toward higher temperatures is more subtle to explain. It is due to the evolution of the distribution of relaxation times during annealing of the PVAc. Generally relaxation times increase during aging. The decay law of the energy during aging generally does not follow one single exponential function, particularly for polymeric glass-former \cite{cangialosi1,cangialosi2}. 

\subsection{Energy uncertainty}
From the experiments 1 and 2, the following integral is calculated:
\begin{equation}
\oint c_{PVAc}(T)dT
\end{equation}
Knowing that for those two liquid-glass-liquid cycles no energy loss occur in the DSC measurement, whatever the scanning rates, the first law states that this cyclic integral must vanish. The result of such an integration is summarized in table II (it is noticed $\varDelta H_{l-g-l}$ like in Ref. \cite{tombari2}). It is of $5.3\times10^{-3}$ J/g for the experiment 1 and $3.1\times10^{-3}$ J/g for the experiment 2. The initial value of the energy is taken equal to zero by convention $H(T_{0})=0$ at the starting value of the equilibrium temperature in the liquid state $T_{0}$. This uncertainty value on the energy is the sum of the uncertainties coming from the calorimeter itself (noise, drift, calibration), our own calibration with Vespel-reference sample, and our data processing  errors. These two values may be compared to those measured by Tropin and co-workers for Polystyrene and PLC glasses (as an exemple the authors found a value of $8\times10^{-3}$ J/g for cyclic integration on the specific heat of polystyrene) \cite{schmelzer1} and to those measured by Tombari and Johari for Polystyrene (they found values of $2\times10^{-4}$ J/g and $3\times10^{-4}$ J/g depending on the scanning rate used) \cite{tombari2}. Our experiments allow us to define a mean uncertainty in enthalpy of $\pm2.1\times10^{-3}$ J/g.  
\\
For the experiment 3, since during the annealing process of 60 h there is no means with the DSC to measure this energy decrease, then the cyclic integration must yield to a positive value in energy that is exactly equal to this annealing value. In other words, this enthalpy loss during the cooling path is inevitably measured by the DSC during heating. We call this value $\varDelta H_{l-g1-g2-l}$ because for these types of experiments (experiments 3, 4 and 5) the PVAc goes in a glassy state by cooling to the aging temperature (called glass1,) and goes then to another (more stabilized) glassy state (called glass2) after annealing. For the ending path of the cooling stage, between the aging temperature and the lower temperature, we suppose that the glass2 has no time to evolve significantly toward a glass3. The values of these annealing enthalpies are given in table 2. We suppose that the uncertainty on these values are of the same order than those directly measured on liquid-glass-liquid cycles.

\subsection{Determination of the fictive temperature}
In Fig. (8), we observe that it is possible to assign a linear baseline for the liquid state and a linear baseline for the glassy state (see the two black dashed lines in Fig. (8)). Let be $C_{l}$ the liquid heat capacity baseline defined on all the temperature range. Let be $C_{g}$ the glass heat capacity baseline defined on the same temperature range. With these two straight lines, we can determine the difference $\varDelta C_{eq}=C_{l}-C_{g}$ and the difference $\varDelta C=C_{PVAc}-C_{g}=C_{i}$ on all the temperature range. Here $C_{i}$ is the configurational heat capacity of the PVAc. It has the same meaning as the configurational heat capacity such as described in the first part of this paper for the TLS system. The normalized heat capacity is defined as the ratio of the two previous heat capacity differences :
\begin{equation}
C^{N}(T)=\frac{\varDelta C(T)}{\varDelta C_{eq}(T_{f})}
\end{equation}
where the denominator $\varDelta C_{eq}$ is taken at the value $T=T_{f}$ the fictive temperature. By means of the Tool equation (Eq.(4)), we calculate the fictive temperature by integration along temperature of the normalized heat capacity:
\begin{equation}
T_{f}(T)=T_{f}(T_{0})+\intop_{T_{0}}^{T}C^{N}(T)dT
\end{equation}
At the starting temperature $T_{0}$, $T_{f}(T_{0})=T_{0}$. Since $C^{N}(T)$ depends also on $T_{f}$, then we calculate a first approximated  $T_{f}(T)$ by integration of a normalized heat capacity with the denominator $\varDelta C_{eq}(T)$ taken at the real temperature. This first approximated $T_{f}(T)$ is then used for having the denominator $\varDelta C_{eq}(T_{f})$ as a function of the fictive temperature in order to be able to integrate a more exact normalized heat capacity. This process is iterated until we obtain a non-moving $T_{f}(T)$ curve on all the temperature range. In Fig. (9), the $T_{f}(T)$ curves are shown for cooling and heating of the experiments 1 and 3. The focus is made on the transition range where the fictive temperature already departs from the temperature. 
\begin{figure} 
\begin{center}
 \includegraphics[width=8.5cm]{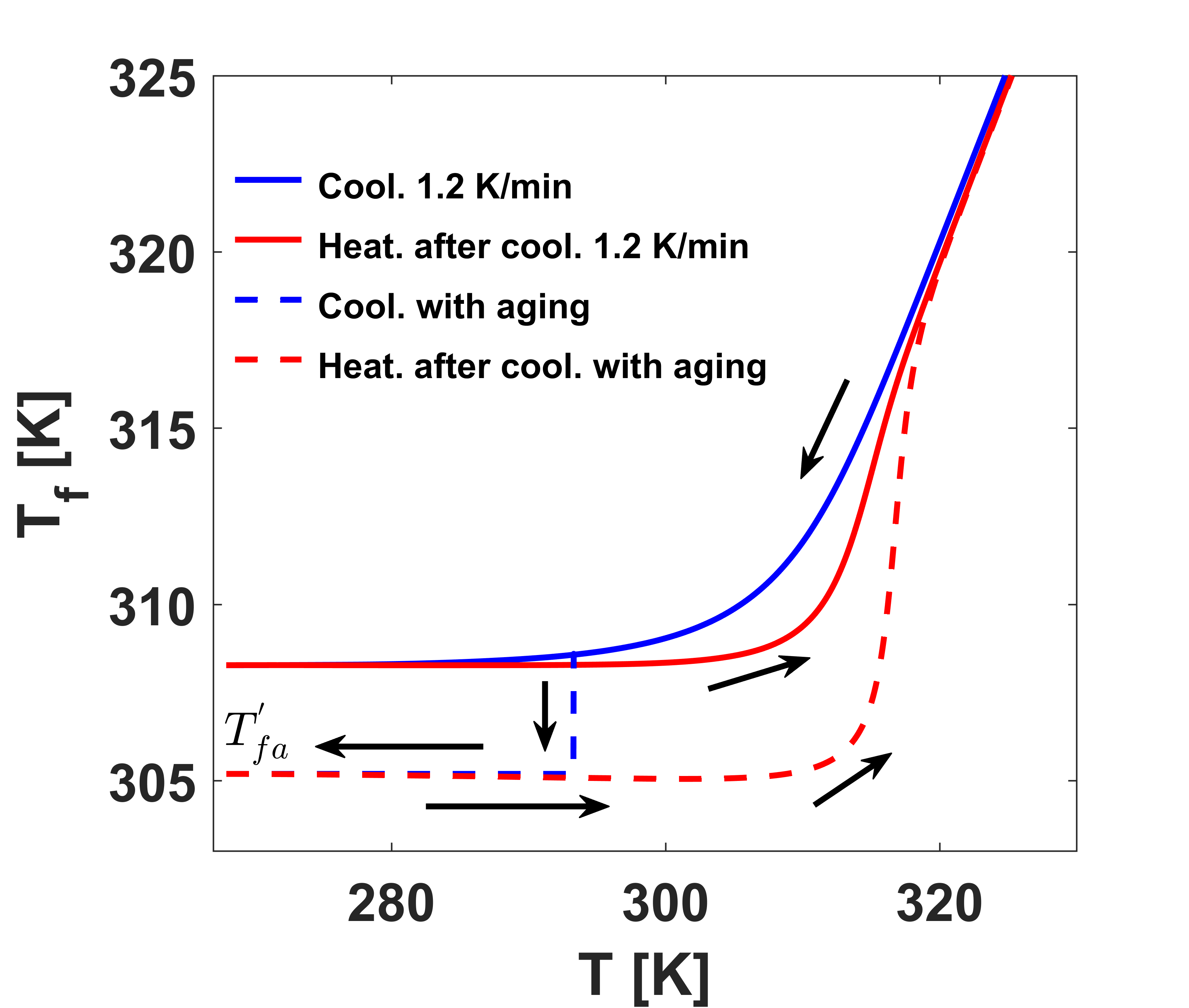}
 \end{center}
 \caption{Fictive temperature of PVAc as a function of temperature for experiment 1 and 3 such as described in table I. The black arrows indicate the sense of the thermodynamic cycle followed by PVAc during the experiments with an aging step at 293.3K for the experiment 3.}
 \label{Fig8}
\end{figure}
The constant values of the fictive temperature reached at low temperature for the two experiments are the limiting fictive temperature $T_{f}^{'}$ of the glassy state. In a way, they represent the energetic state of the glass obtained  after a given thermal history. The cooling at 1.2 K/min and the successive heating at the same rate of experiment 1 is seen as solid blue line and solid red line respectively. For the aging experiment, the $T_{f}(T)$ curves are represented by blue dashed line for the cooling (with aging as vertical line) and red dashed line for the heating. For this experiment, contrary to the TLS model, it is not possible to extract the fictive temperature along time. We don't know the relaxation law of the fictive temperature at constant temperature. In the experiment 3, since it is not possible to follow the trajectory of the fictive temperature along time, then we have however the means to determine the final value of the fictive temperature at the end of the annealing process ($T_{fa}^{'}$ in Fig. (9)). For that purpose, we assume that at the end of the cycle, in the liquid state the ending value of the fictive temperature is equal to the initial experimental temperature. Therefore, by inverse integration of the normalized heat capacity along the heating path, we can come back to the value of the fictive temperature reached at the end of aging. We thus have two well-defined values of the fictive temperature, one for the glass1 (just before the aging), and the other for the glass2 (just after the aging). A vertical line is then traced (see blue dashed line in Fig.(9))between these two boundaries values of the aging process. The second assumption we used, is to consider that between the aging temperature of 293.30 K and the final low temperature of 267.95 K, the fictive temperature remains constant. For an experiment without annealing step, the decrease in the fictive temperature under the same temperature range is of only around 0.3 K. This is negligible as compared to the few kelvins of decrease during aging (see Fig. (9)). Moreover, it is reasonable to consider that after a period of annealing, the decrease of the fictive temperature is even less than 0.3 K, because the glass2 is already stabilized. This small cooling path in the glass2 state is barely seen as a horizontal blue dashed line in Fig. (9).

\subsection{Determination of the entropy production rate}
Like in the first part of this paper, we have now at our disposal all the necessary parameters to determine the entropy production rate for each experiment, either during cooling or during heating, by means of Eq(5):
\begin{equation}
\sigma_{i}=\frac{d_{i}S}{dt}=\left(\frac{C_{i}}{T_{f}}-\frac{C_{i}}{T}\right)\frac{dT}{dt}
\end{equation}
The result is shown in Fig.(10) for the cooling of the experiments 1 and 2 (blue solid and dashed lines respectively), and for the three heating of the experiments 1 (red solid line), experiment 2 (red dashed line), and experiment 3 (red dotdashed line).
\begin{figure} 
\begin{center}
 \includegraphics[width=8.5cm]{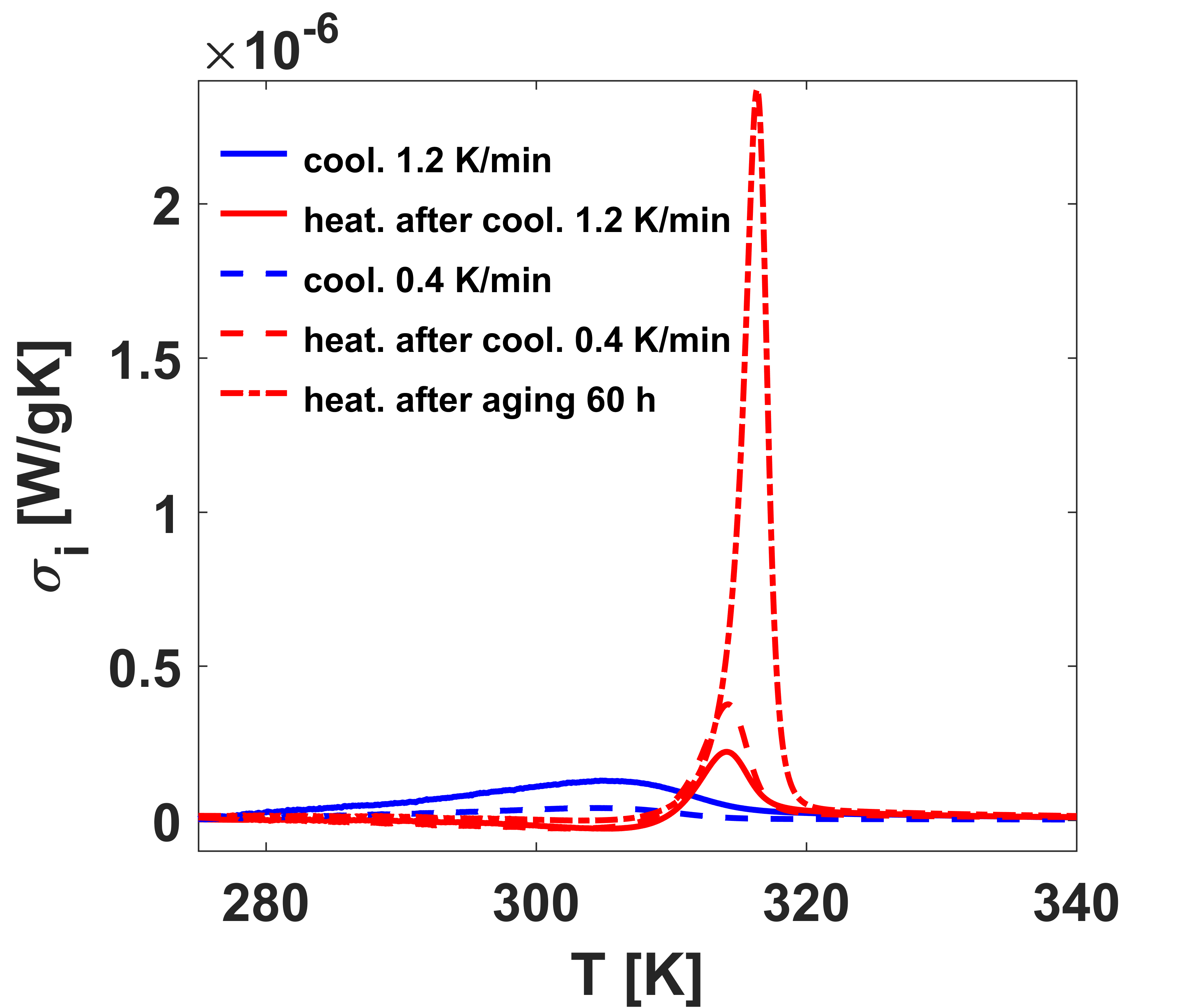}
 \end{center}
 \caption{Entropy production rate of PVAc as a function of temperature for the three experiments 1, 2 and 3 such as described in table I. The cooling path for the experiment 3 with aging is not represented.}
 \label{Fig10}
\end{figure}
For cooling, there is one small but broad peak of entropy production. The higher the cooling rate, the higher the entropy production peak. For heating, contrary to simulations with TLS, there is only one well-visible entropy production peak. The smaller the previous cooling rate, the higher the peak during heating. The peak of the heating after aging is particularly higher and broader than peaks issued from simple cooling. This aspect is discussed in a next section. In focusing in the low temperature range, we see the limit of our data processing since the entropy production rate is slightly negative for the three experiments at the starting of heating, which is unexpected. A new definition of the baselines specific heat, particularly in the glassy state, might avoid this problem, but we have decided in this work to take the same specific heat baselines for all the data processing, the enthalpy integration, and the fictive temperature determination. It is however easy to imagine the appearance of one first small entropy production peak in this low temperature range with a glass-specific heat baseline which is not a straight line. 
\\
To the best of our knowledge, it is the first time that the entropy production rate is determined as a function of temperature from a calorimetric DSC-experiment. Although the presence of such peaks during cooling and heating is already the demonstration of the Clausius theorem, we would like to validate it under its usual integrated shape (see Eq. (10) above) and discuss more deeply the case of aging.

\subsection{Verification of the Clausius theorem}
\subsubsection{Liquid-glass-liquid cycles}
Firstly, we would like to calculate the entropy produced along the liquid-glass-liquid cycles, \textit{i.e.} for the experiments 1 and 2. In a first step, we calculate the following integral:

\begin{equation}
\varDelta S_{l-g-l}=\oint\frac{C_{i}}{T_{f}}dT
\end{equation}

The final result of such cyclic integration is presented in table II. The resulting values are close to zero such as expected. The errors are calculated as the mean value of these two values like for energy uncertainty. We find $\pm1.2\times10^{-5}$ J/gK as combined errors of our entropy determination. In a second step, we calculate the usual calorimetric integral:

\begin{equation}
\varDelta S_{l-g-l}^{heat}=\oint\frac{C_{i}}{T}dT
\end{equation}

The results are also shown in table II. These values of the entropy calculated from the configurational heat capacity divided by temperature along the cycle are lower (and negative) than their corresponding values calculated with the integral (16) above. The difference at the end of integration is about ten times more significant than the entropy uncertainty. This confirms the validity of the Clausius theorem with missing net entropy production term in the latter integral. For the sake of clarity, we have summarized these results in Fig. (11).
\begin{figure} 
\begin{center}
 \includegraphics[width=8.5cm]{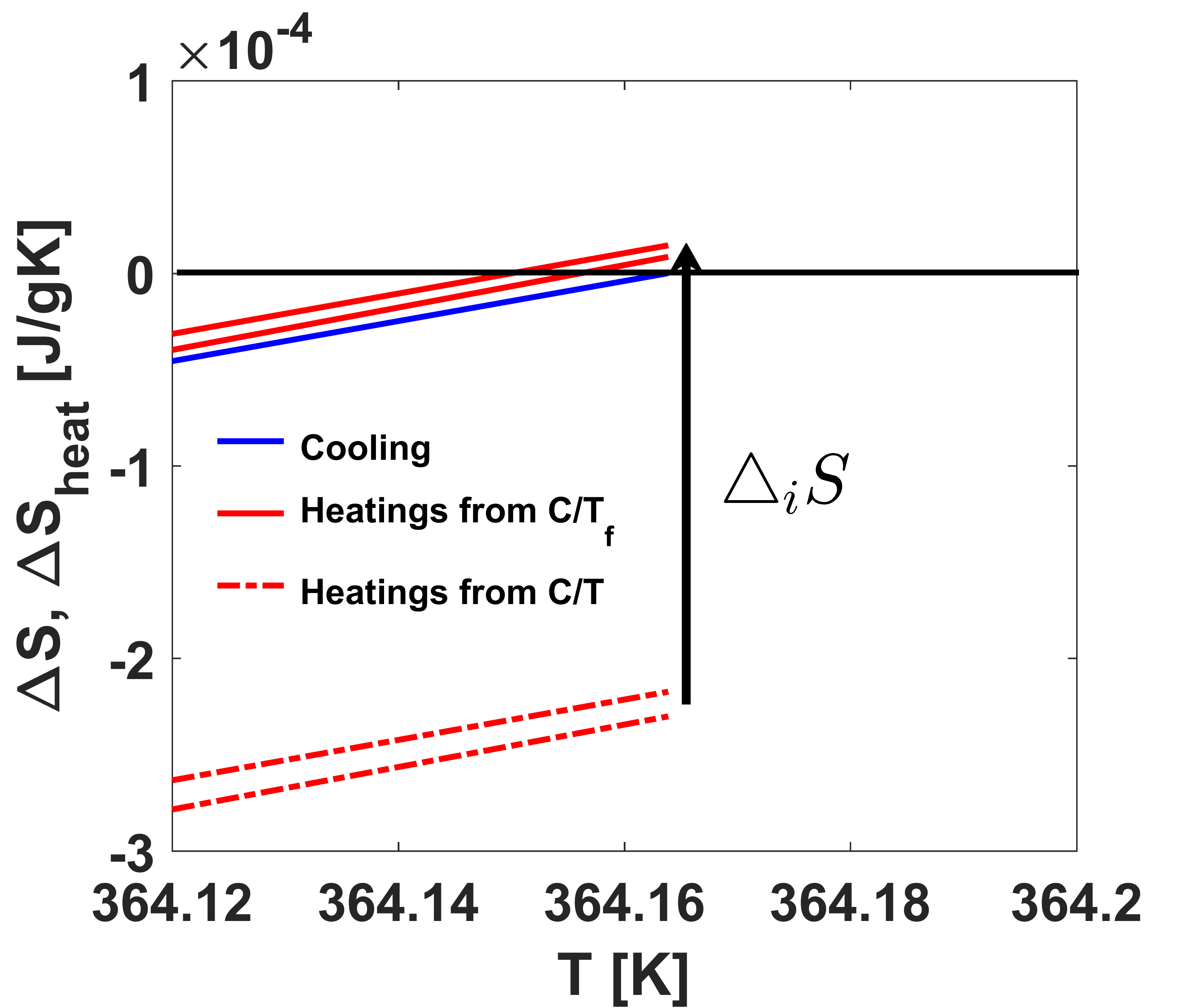}
 \end{center}
 \caption{Configurational entropy of PVAc obtained by integration of $C_{i}/T$ and $C_{i}/T_{f}$ as a function of temperature during cooling and the successive heating for the experiments 1 and 2 such as described in table I. The figure shows a magnification of the configurational entropy in the temperature range around the starting and final temperature.}
 \label{Fig11}
\end{figure}
The initial value of the integrals above are taken equal to zero by convention. The horizontal black line shows the level zero for a sake of clarity. The blue solid line shows the first starting values for all the integrals with zero as starting point. The two red solid lines show the final values of the integral (16) for the experiment 1 and 2, while the two red dotdashed lines show the final values of the integral (17)(see table II). As clearly shown in Fig. (11), the difference in the calculation of the two integrals above is about ten times higher than the dispersion around their final values for the experiments 1 and 2. The Clausius theorem is verified in this case. For a last verification, we may verify for each experiment that the difference $\varDelta_{i}S$ obtained by the difference between (16) and (17) is exactly the same as that obtained by integration along the cycle of the entropy production rates shown in Fig. (10). 

\subsubsection{Liquid-glass1-glass2-liquid cycles}
This section deals with aging experiments. In this case, we have to consider particular thermodynamic cycles of the type liquid-glass1-glass2-liquid, for which the first glass in the cycle (glass1) is obtained by cooling to the aging temperature, and the second one (glass2) is obtained after the aging time at the same temperature. This is the case of experiments 3, 4 and 5. The experiments 4 and 5 are identical to the experiment 3, except for the aging time which is of about 20 h for the experiment 4, and about 6 h for the experiment 5. As already discussed, it is not possible for such experiments to directly measure the enthalpy and entropy decrease during aging. It is also not possible to directly measure the entropy produced during such relaxation process. While this decrease in entropy and enthalpy are recovered by the measurement of the specific heat during the successive heating scan, for the entropy production, it is not possible to recover it. Indeed, this entropy is by nature irreversibly produced and irremediably lost. However, it is possible to determine this contribution using the same approach than that we used in the determination of the fictive temperature during aging (see the vertical blue dashed line in Fig.(9)). For the three aging experiments 3, 4 and 5, the fictive temperature of the glass1 at 293.3 K is obtained by integration of the normalized heat capacity during cooling. Since the cooling rate is the same, this glass1 fictive temperature is the same for the three experiments. After the aging process, the final values of the fictive temperatures are calculated by inverse integration along the heating paths assuming that the ending values of the fictive temperatures must equal the initial equilibrium liquid temperature (364.16 K). Fictive temperatures calculated like this are obviously different for the three experiments, and they are each associated to a particular glassy state (glass2). The second assumption is that these particular values do not vary when the end of cooling is carried up (between the aging temperature and the lowest temperature of 267.95 K). By this way, we have access to the entropy production during aging even if the process of relaxation is not known. For each experiment, these specific fictive temperatures (just before and just after the aging) are used to determine the entropy produced during this isothermal process. For an isothermal process, the expression (15) is not valid anymore, since it is adapted only for stages where the temperature varies (cooling or heating). Using Tool's equation, we use the other expression for the entropy production rate in terms of fictive temperature:

\begin{equation}
\sigma_{i}=\frac{d_{i}S}{dt}=\left(\frac{C_{i}^{eq}}{T_{f}}-\frac{C_{i}^{eq}}{T}\right)\frac{dT_{f}}{dt}
\end{equation}

where $C_{i}^{eq}$ has to be taken at the value of the fictive temperature such as already explained. The integration of the expression above is processed over the fictive temperature along all the cycle liquid-glass1-glass2-liquid. The results of such an integration is shown in Fig. (12) as a function of temperature for the three experiments 3, 4 and 5. It is also shown along the cycle of the experiment 1 without aging for comparison. 
\begin{figure} 
\begin{center}
 \includegraphics[width=8.5cm]{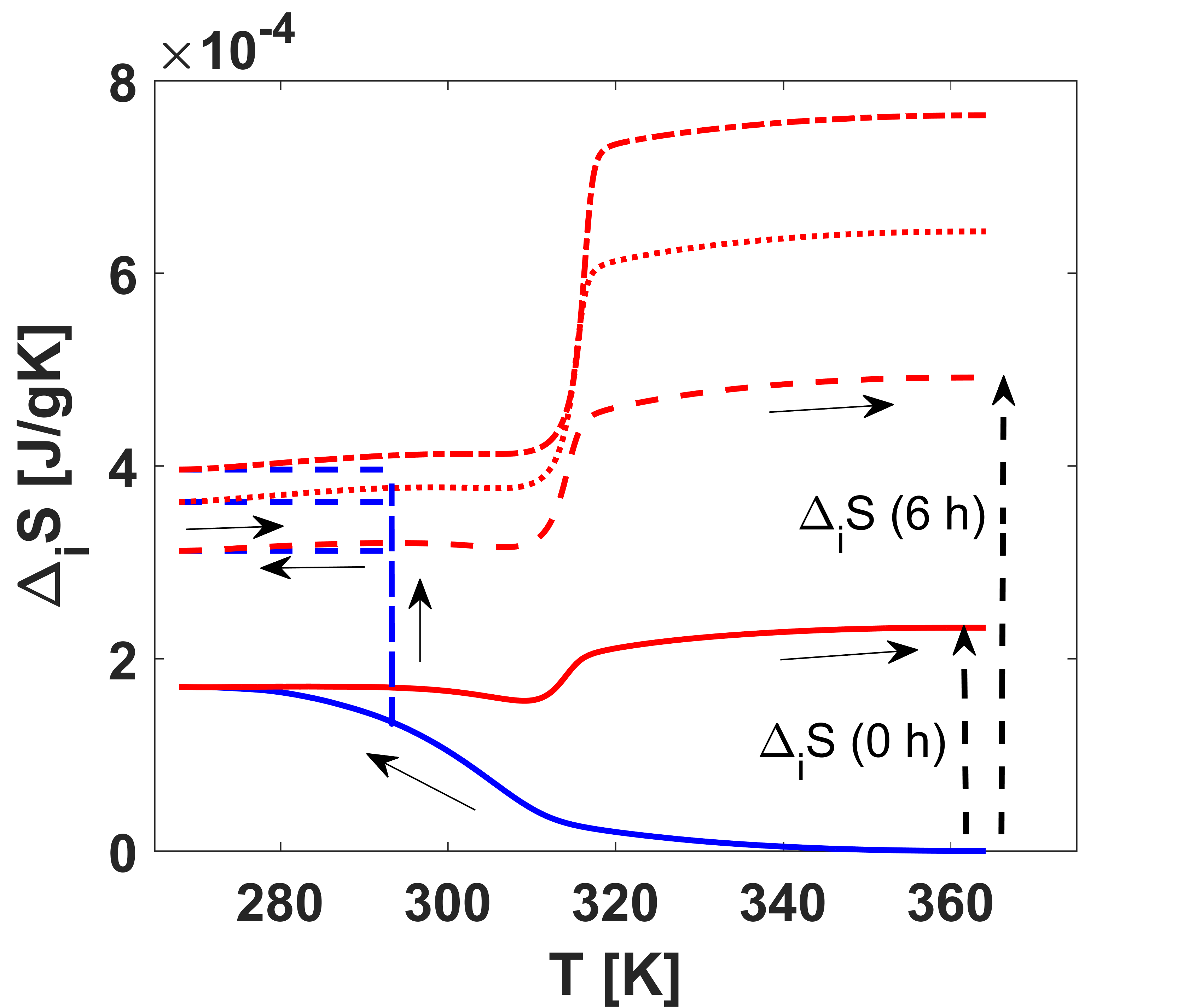}
 \end{center}
 \caption{Entropy production contributions for PVAc as a function of temperature during different parts of the thermodynamic cycles for the experiment 1 during cooling and the successive heating, and for the experiments 3, 4 and 5 during cooling, during annealing, during the rest of cooling, and during the successive heating such as described in table I. The black arrows indicate the sens of progression of the thermodynamic cycle for the experiment 5 with 6 h of aging at 293.3K. }
 \label{Fig12}
\end{figure}
The blue solid line is the entropy produced upon cooling at - 1.2 K/min. The blue dashed lines are the entropy productions during the cooling but with aging time varying between 6 h (experiment 5), 20 h (experiment 4) and 60 h (experiment 3). Aging is seen as a vertical line stopping at different levels depending on the aging time, and continuing as horizontal blue dashed lines till the lowest temperature. The red solid line is the entropy produced during heating at 1.2 K/min for the experiment without aging. The entropies produced during heating at 1.2 K/min after the aging time of 6 h, 20 h and 60 h, are seen as red dashed line, red doubledotted line and red dotdashed line respectively. For the sake of clarity, arrows indicate the sens of entropy production for the experiment 1 without aging, and for the experiment 5 (aging 6h) only. For such experiments 1 and 5, we may observe a slight decrease of the entropy just before the jump due to the unfreezing of the PVAc, which is unexpected. This indicates the limit of our data processing protocol due to the combination of different errors such as already discussed (particularly the choice of a glass specific baseline that we have chosen as a unique straight line for all the experiments). Nevertheless, for all the experiments, we see that the Clausius theorem is fulfilled, with final values of the positive entropy produced over all the considered thermodynamic cycles well above the total uncertainty of the experiments. In table II, we indicate the values of the entropy produced at different stages of the cycles. In particular, in table II all the $\Delta_{i}S$ are integration of $\sigma_{i}$ over the fictive temperature (cf. Eq. (18)) between the liquid and glass with the subscript $l-g$ for cooling experiment, between the glass and liquid with subscript $g-l$ for heating experiment, between liquid and glass1 with the subscript $l-g1$ for the path of the aging experiment between the liquid and the glass reached before annealing at the aging temperature, between glass1 and glass2 with the subscript $g1-g2$ for the path of the aging experiment during annealing, and between glass2 and liquid with the subscript $g2-l$ for the aging experiment during the final heating. The net entropy produced along all the cycle of the aging experiment have the subscript $l-g1-g2-l$. Finally, values of the configurational entropy variation during the cooling (or identically during the heating but with an opposite sign), noted $\Delta S_{l-g}$ and $\Delta S_{g-l}$, are also calculated in table II for comparison. All these data are commented in the next discussion.
\\
Before commenting these data, we would like to stress a technical point of interest. In table II, the values of $\Delta H_{l-g1-g2-l}$, $\Delta S_{l-g1-g2-l}$, and $\Delta S_{l-g1-g2-l}^{heat}$ are different from zero because we used the integration over the temperature, which does not take into account the annealing process at the constant aging temperature. The values given in the table II correspond then to the annealing enthalpies and entropies of the experiments 3, 4 and 5. If we have proceeded by using integration over the fictive temperature, like we did for the determination of the entropy production contributions in table II, we then have found a value very close to zero (up to the experimental inaccuracy) for  $\Delta H_{l-g1-g2-l}$ and $\Delta S_{l-g1-g2-l}$, and slightly below zero (due to the lack of entropy production contribution) in  $\Delta S_{l-g1-g2-l}^{heat}$. In appendix A, we show the configurational entropies   over all the thermodynamic cycle for the experiment 3 (60 h of aging) when integration is made over temperature and over fictive temperature.  

\begin{table*}
\begin{tabular}{|c|c|c|c|c|c|}
\hline 
 & Exp. 1 & Exp. 2 & Exp. 3 & Exp. 4 & Exp. 5\tabularnewline
\hline 
\hline 
$\Delta H_{l-g-l}$ (J/g) & $5.3\times10^{-3}$ & $3.1\times10^{-3}$ & - & - & -\tabularnewline
\hline 
$\Delta S_{l-g-l}$ (J/gK) & $1.5\times10^{-5}$ & $8.5\times10^{-6}$ & - & - & -\tabularnewline
\hline 
$\Delta S_{l-g-l}^{heat}$(J/gK) & $-2.2\times10^{-4}$ & $-2.3\times10^{-4}$ & - & - & -\tabularnewline
\hline 
$\Delta_{i}S_{l-g}$(J/gK) & $1.7\times10^{-4}$ & $1.5\times10^{-4}$ & - & - & -\tabularnewline
\hline 
$\Delta_{i}S_{g-l}$(J/gK) & $6.1\times10^{-5}$ & $5\times10^{-5}$ & - & - & -\tabularnewline
\hline 
$\Delta_{i}S_{l-g-l}$(J/gK) & $2.3\times10^{-4}$ & $2\times10^{-4}$ & - & - & -\tabularnewline
\hline 
$\Delta H_{l-g1-g2-l}$ (J/g) & - & - & $1.6$ & $1.3$ & $1$\tabularnewline
\hline 
$\Delta S_{l-g1-g2-l}$ (J/gK) & - & - & $5.2\times10^{-3}$ & $4.4\times10^{-3}$ & $3.2\times10^{-3}$\tabularnewline
\hline 
$\Delta S_{l-g1-g2-l}^{heat}$(J/gK) & - & - & $4.6\times10^{-3}$ & $3.9\times10^{-3}$ & $2.9\times10^{-3}$\tabularnewline
\hline 
$\Delta_{i}S_{l-g1}$(J/gK) & - & - & $1.3\times10^{-4}$ & $1.3\times10^{-4}$ & $1.3\times10^{-4}$\tabularnewline
\hline 
$\Delta_{i}S_{g1-g2}$(J/gK) & - & - & $2.6\times10^{-4}$ & $2.3\times10^{-4}$ & $1.8\times10^{-4}$\tabularnewline
\hline 
$\Delta_{i}S_{g2-l}$(J/gK) & - & - & $3.7\times10^{-4}$ & $2.8\times10^{-4}$ & $1.8\times10^{-4}$\tabularnewline
\hline 
$\Delta_{i}S_{l-g1-g2-l}$(J/gK) & - & - & $7.6\times10^{-4}$ & $6.4\times10^{-4}$ & $4.9\times10^{-4}$\tabularnewline
\hline 
$\Delta S_{l-g}$(J/gK)  & $-7.46\times10^{-2}$ & $-7.8\times10^{-2}$ &$-8\times10^{-2}$  &  & \tabularnewline
\hline 
$\Delta S_{g-l}$(J/gK) & $7.46\times10^{-2}$ & $7.8\times10^{-2}$ & $8\times10^{-2}$ &  & \tabularnewline
\hline 
\end{tabular}
\caption{Enthalpies and entropies measured on total and partial thermodynamic cycles. The mean combined error for the enthalpies is of $\pm2.1\times10^{-3}$ J/g and the mean combined error for the entropies is of $\pm1.2\times10^{-5}$ J/g.K }
\end{table*}

\section{Discussion and conclusion}  
The TLS statistical model, when it is coupled to a master kinetic equation, is able to reproduce the primary features of the glass transition. In particular, the configurational heat capacity and the fictive temperature are calculable within this framework. With a simple thermal model, by means of these two parameters, we can calculate the entropy production rate along a thermodynamic path followed by the TLS. It is possible to verify the validity of the Clausius theorem under this framework. An important point is that the total entropy of the system may be obtained by integration over temperature of the configurational heat capacity divided by the fictive temperature. If we integrate the configurational heat capacity divided by the classical temperature, like it is always performed in equilibrium thermodynamics, we lose the positive entropy production term, and the final entropy calculated on the entire thermodynamic cycle is smaller than the initial one. This is due to the fact that the passage from an equilibrium TLS to a kinetically arrested TLS (and \textit{vice versa}) is by essence an irreversible process whatever the thermal-history of the system. At this level, a point of significant interest is the importance of the parameter "fictive temperature" in the determination of the total entropy of the system. Without the calculation of the fictive temperature (owing to the integration of the normalized heat capacity), it is not possible to access to the exact value of the glass residual entropy. Moreover, during annealing steps, this is the knowledge of the fictive temperature at the aging temperature which allows us to determine the production of entropy during such isothermal processes. A calorimeter able to measure also the heat-flux between the sample and the surroundings under isothermal condition (for exemple with power compensation technique) might in principle provide directly this isothermal entropy production at the same time than the annealing entropy. As a perspective for this discussion on the fictive temperature, in the expression (9) of the configurational entropy variation, the statistical average of $-k_{B}lnp_{i}$ on the two states of the TLS already contains intrinsically the fictive temperature parameter. How can we extract this macroscopic parameter from statistical physics might be a question of interest for the study of the glass transition at the microscopic scale.
\\ 
Inversely, all these considerations may be applied to real glassy systems. By means of DSC-measurements of the heat capacity of a glass (here the PVAc polymeric glass-former), it is possible to calculate the fictive temperatures. Therefore, it is possible to determine the rate of production of entropy generated inside the PVAc-system along different phases of the liquid-glass-liquid transition path. This is what we have done for the first time experimentally. We have experimentally observed one broad and small peak during cooling, and one highly resolved peak during heating for all the different types of experiments considered in this work. Moreover, the combined errors due to the DSC experiments and the data processing used in this study are about ten times smaller than the positive entropy determined by integration on the entire considered thermodynamic cycle. For each experiment, we have verified that the integrated form of the Clausius theorem is fulfilled, with values going from  $2.3\times10^{-4}$ J/gK for the smallest one (experiment 1) to $7.6\times10^{-4}$ J/gK for the highest value determined for the experiment with 60 h of aging (experiment 3). This is for this latter experiment than the entropy created over the cycle is the highest. However, if we compare this value with the annealing entropy ($5.2\times10^{-3}$ J/gK) determined from the same experiment, this is only about 15 \%. And, if we compare it to the configurational entropy between the liquid and the glassy state ($8\times10^{-2}$ J/gK), it is only about 1 \%. This ratio is only about 0.3 \% for the experiment 1 without aging. Let us emphasize here that such results have been possibly recorded owing to a fine and accurate treatment of the data, and an important care has been taken into account in the calibration of the DSC. A reference material (Vespel) has been used in a first step in order to calibrate the calorimeter at different scanning rates, either for cooling or for heating. If we compare our results with those obtained by Tropin and colleagues, we observe that these authors validate the Clausius theorem with a value of $-7\times10^{-5}(\pm2\times10^{-5})$ J/gK from the DSC calculation of entropy on polystyrene, and $-4\times10^{-4}(\pm1\times10^{-4})$ J/gK for PCl at 300 K/s and $-1.5\times10^{-4}(\pm1\times10^{-4})$ J/gK for PCL at 50,000 K/s \cite{schmelzer1}. The estimated error is 57 \% of the cyclic integral for polystyrene, and  50 \% for PCl at 300 K/s, and 133 \% for PLC at 50,000 K/s. For a sake of comparison, our calculations for the same integrals are  $-2.2\times10^{-4}(\pm1.2\times10^{-5})$ J/gK for experiment 1, and $-2.3\times10^{-4}(\pm1.2\times10^{-5})$ J/gK for experiment 2. Our estimated error is only about 11 \% of the cyclic integral. With an highly sensitive home-made calorimeter, Tombari and Johari obtained values of errors of about 8\% of the cyclic integral with a value of  $4.8\times10^{-5}$ J/gK \cite{tombari2}. Surprisingly, they obtained positive values, while the values of Tropin and co-workers, and ours, are negative such as predicted by the Clausius theorem. In accordance with the cited authors, we arrive at the conclusion that the entropy dissipated as uncompensated heat at the molecular scale is generally small as compared to the corresponding entropic configurational change. For the experiment with equivalent cooling and heating rate (1.2 K/min), there is only about 36 \% of the total entropy produced on the cycle which is dissipated during the glass to liquid structural recovery ($\Delta_{i}S_{g-l}$) as compared to the entropy produced during freezing ($\Delta_{i}S_{l-g}$). For all the other experiments, the entropy produced during structural recovery (unfreezing) is higher than the entropy produced during freezing of the molecular movements. Unfreezing of a slowly cooled, or of a aged glass, produces more entropy than during the slow cooling, but also generally it produces more entropy than during annealing ($\Delta_{i}S_{g1-g2}$). In the experiments presented here, this is for the 6 h annealed glass that the entropy produced during annealing is equivalent to the entropy produced during unfreezing ($1.8\times10^{-4}$ J/gK). For the two other annealed PVAc glasses, the entropies produced are higher during unfreezing. The ratio of the entropy produced during annealing on the entropy produced during unfreezing is of about 82 \% for the 20 h annealing, and 70 \% for the 60 h annealing. As can be extracted from these three aging experiments, the entropy production increases with the aging time, whatever the process involved in this production, either during annealing (\textit{i.e} during the decrease in the landscape), or during unfreezing (during the structural recovery of the liquid state). This positive entropy contributions seem to tend towards an asymptotic value as the aging time increases, but we need more aging experiments to conclude. The annealing energies and entropies associated to these processes are also higher for higher aging times, with also a tendency to an asymptotic behaviour. 
\\
Finally, the understanding of such phenomenon on a fundamental point of view may be of pertinence and utility for the calibration of a calorimeter and the determination of liquid and glass baselines. Let us consider only thermodynamic cycles consisting in starting from an equilibrium state and in coming back to the same initial state after a cooling followed by a heating. In this case, we may assume without any doubt that the three following thermodynamic relations are true:
\\
- the final value of the fictive temperature has to be equal to the same initial value, after the integration on the cycle of the normalized heat capacity :

\begin{equation}
\oint dT_{f}=0
\end{equation}

- the final value of the energy calculated by integration on the cycle of the heat capacity  has to be equal to the initial value:

\begin{equation}
\oint C_{i}^{eq}(T_{f})dT_{f}=0
\end{equation}

- the final value of the total entropy calculated by integration on the cycle of the heat capacity divided by the fictive temperature has to be equal to the initial value:

\begin{equation}
\oint C_{i}^{eq}(T_{f})d\ln T_{f}=0
\end{equation}

For thermodynamic cycles containing stages of annealing, it is more appropriate to proceed to the integration of the variables over the fictive temperature if the change of this parameter can be known with a procedure similar to what we explained in this work (see also the Appendix).
\\
As a consequence, with a chosen glassy material, the DSC may be calibrated along a cycle (with a chosen thermal history) defined by a liquid-glass-liquid path, or other thermodynamic paths with the only constraint that it has to start in the equilibrium liquid state. The DSC may be calibrated  with liquid and glass specific heat baselines fulfilling these three previous equalities.

\section{acknowledgments}
We would like to gratefully thank M. Peyrard for his help, his judicious comments and fruitful discussions we have with him about this work. We would also thank H. Guillou for the numerous daily discussions on non-equilibrium thermodynamics. 

\section{Appendix} 
This short appendix shows in Fig.(13) the difference between the configurational entropy calculated by integration over the temperature, and the configurational entropy calculated by integration over the fictive temperature for an experiment containing a stage of annealing at a constant temperature (here experiment 3 with 60 h of aging).

\begin{figure} 
\begin{center}
 \includegraphics[width=8.5cm]{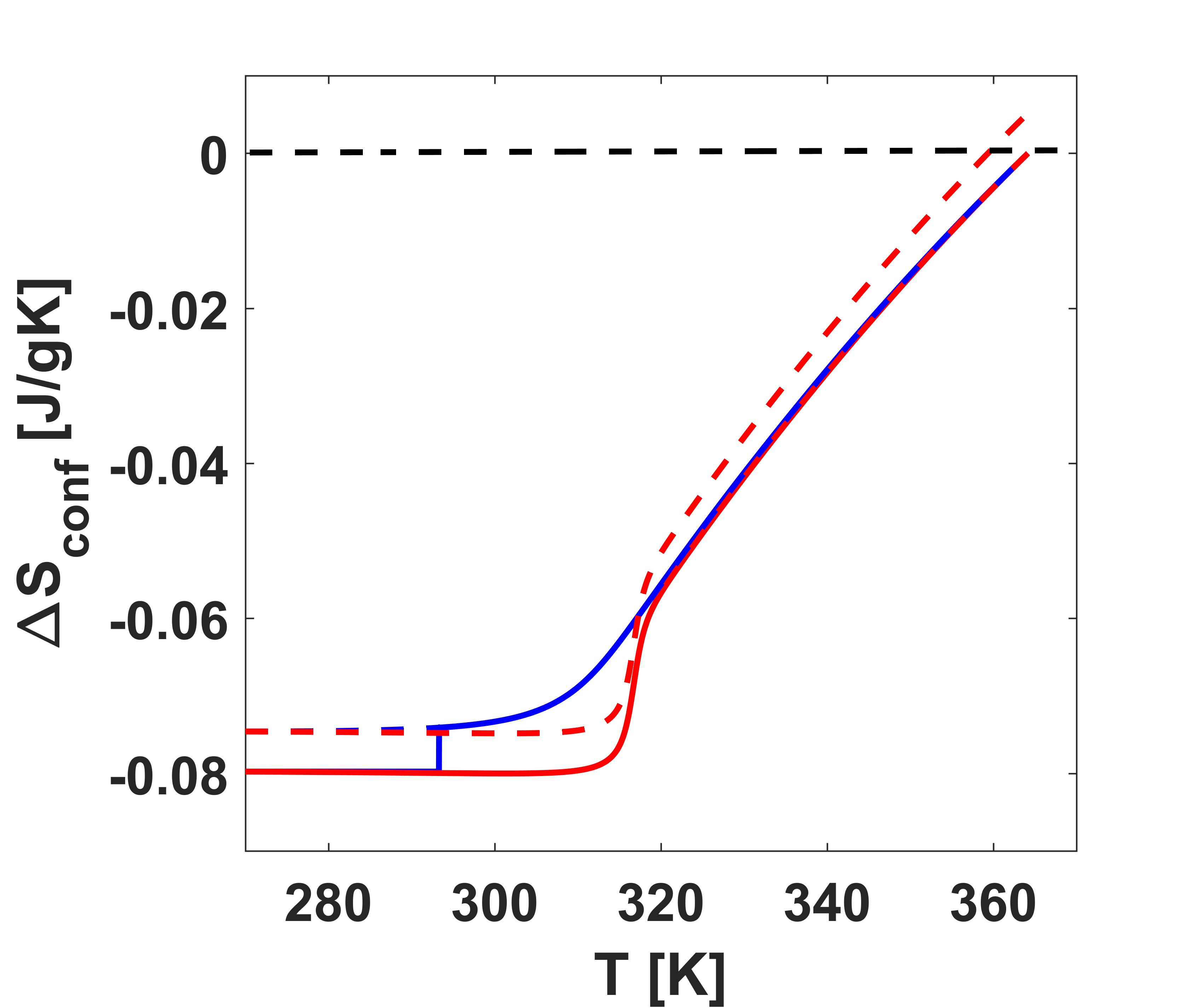}
 \end{center}
 \caption{Configurational entropy for PVAc as a function of temperature for the experiment 3 with 60 h of aging such as described in table I, obtained by integration over temperature (dashed lines)and by integration over the fictive temperature (solid lines) of the configurational heat capacity divided by the fictive temperature.The horizontal black dashed line indicate the zero level of entropy (starting point).}
 \label{Fig13}
\end{figure}

The starting value of entropy is taken equal to zero by convention. The cooling entropy obtained by integration over temperature is the blue dashed line while the cooling entropy obtained by integration over the fictive temperature is the blue solid line. The latter shows an isothermal jump at the aging temperature since the fictive temperature undergoes an isothermal relaxation at the aging temperature. The $\Delta S_{conf}$ at this aging temperature is the configurational entropy decrease during annealing of the PVAc at 293.3K. The red dashed line is the heating entropy determined by integration over temperature while the red solid line is the heating entropy determined by integration over the fictive temperature. The final value of the former arrives well above zero, while the final value of the latter arrives exactly to zero. 
In other words, by integration over the temperature, the annealing step occurring at constant temperature cannot be considered, and the final value provides this annealing configurational entropy. In considering that the final value of the fictive temperature must be equal to the initial one (and equal to the classical temperature), then in using an inverse integration on the last heating path, the step in the fictive temperature due to annealing may be estimated. Therefore, by integration over the fictive temperature, this annealing step is taken into account and the final value of the configurational entropy is exactly equal to the initial one.

\bibliographystyle{plain}

\begin{thebibliography}{}

\end{thebibliography}


\begin{thebibliography}{18}

\bibitem{prigo4}
I.~Prigogine and R.~Defay,
\newblock {\em Chemical Thermodynamics}.
\newblock (Longmans, 1954).

\bibitem{davies}
R.~O. Davies and G.~O. Jones,
\newblock {\em Advanc. Phys. (Phil. Mag. Suppl.)}, \textbf{2}, 370 (1953).

\bibitem{bestul}
A.~B. Bestul and S.~S. Chang.
\newblock {\em J. Chem. Phys.}, \textbf{43}, 4532 (1965).

\bibitem{goldstein2}
M.~Goldstein.
\newblock {\em J. Chem. Phys.}, \textbf{64}, 4767 (1976).

\bibitem{gutzow3}
J.~Möller, I.~Gutzow, and J.~W.~P. Schmelzer,
\newblock {\em J. Chem. Phys.}, \textbf{125}, 094505 (2006).

\bibitem{schmelzer1}
T.~V. Tropin, J.~W.~P. Schmelzer, and C.~Schick.
\newblock {\em J. Non-Cryst. Solids}, \textbf{357}, 1302 (2011).

\bibitem{tombari2}
E. Tombari, and G.~P. Johari,
\newblock {\em J. Chem. Phys}, \textbf{141}, 074502 (2014).

\bibitem{bisquert1}
J.~Bisquert.
\newblock {\em Am. J. Phys.}, \textbf{73}, 735 (2005).

\bibitem{aquino}
G.~Aquino, A.~Allahverdyan, and T.~M. Nieuwenhuizen,
\newblock {\em Physical Review Letter}, \textbf{101}, 015901 (2008).

\bibitem{takada}
A.~Takada, R.~Conradt, and P.~Richet.
\newblock {\em J. Non-Cryst. Solids}, \textbf{360}, 13 (2013).

\bibitem{tool1}
A.~Q. Tool,
\newblock {\em J. res. Natl. Bur. Stand.}, \textbf{34}, 199 (1945).

\bibitem{tool2}
A.~Q. Tool.
\newblock {\em J. Am. Ceram. Soc.}, \textbf{29}, 240 (1946).

\bibitem{garden1}
J.-L. Garden, J.~Richard, and H.~Guillou,
\newblock {\em J. Chem. Phys.}, \textbf{129}, 044508 (2008).

\bibitem{bridgman}
P.~W. Bridgman,
\newblock {\em Rev. Mod. Phys.}, \textbf{22}, 56 (1950).

\bibitem{garden6}
J.-L. Garden, H.~Guillou, J.~Richard, and L.~Wondraczek,
\newblock {\em J. Chem. Phys}, \textbf{137}, 024505 (2012).

\bibitem{moy4}
H.~Sasabe and C.~T. Moynihan.
\newblock {\em J. Polym. Sci.: Polym. Phys.}, \textbf{16}, 1447 (1978).

\bibitem{cangialosi1}
A.~Alegria D.~Cangialosi, V. M.~Boucher and J.~Colmenero.
\newblock {\em Phys. Rev. Lett.}, \textbf{111}, 095701 (2013).

\bibitem{cangialosi2}
D.~Cangialosi.
\newblock {\em J. Phys.: Condens. Matter}, \textbf{26} 153101 (2014).


\end{thebibliography}

\end{document}